\documentclass{aa}  
\usepackage{graphicx,lscape,epsfig,natbib}
\usepackage{txfonts}
\begin{document}
   \title{The multi-phase gaseous halos of star forming late-type galaxies\thanks{Based on observations obtained with XMM-Newton, an ESA science mission with instruments and contributions directly funded by ESA member states and NASA.}}

   \subtitle{II. Statistical analysis of key parameters}

   \author{R. T\"ullmann
          \inst{1}
          \and
          D. Breitschwerdt\inst{2}
          \and
          J. Rossa\inst{3}
          \and
          W. Pietsch\inst{4}
          \and
          R.-J. Dettmar\inst{1}
          }


  \institute{Astronomisches Institut, Ruhr-Universit\"at Bochum,
              D-44780 Bochum, Germany\\
              \email{tullmann@astro.rub.de,dettmar@astro.rub.de}  
          \and
              Institut f\"ur Astronomie, T\"urkenschanzstrasse 17,
              A-1180 Wien, Austria\\ 
              \email{breitschwerdt@astro.univie.ac.at}
          \and
              Space Telescope Science Institute, 3700 San Martin Drive,  
              Baltimore, MD 21218, U.S.A.\\ 
              \email{jrossa@stsci.edu}
          \and 
              Max-Planck Institut f\"ur extraterrestrische Physik,
              Giessenbachstrasse, D-85748 Garching, Germany\\
              \email{wnp@mpe.mpg.de}
              }

 \date{Received December 21, 2005; accepted April 26, 2006}

\abstract
{In a previous paper (Paper\,I) we showed that multi-phase gaseous halos of late-type spiral galaxies, detected in the radio continuum, in H$\alpha$, and in X-rays, are remarkably well correlated regarding their morphology and spatial extent.}
{In this work we present new results from a statistical analysis in order to specify and quantify these phenomenological relations.}
{This is accomplished by investigating soft X-ray (0.3\,--\,2.0\,keV) luminosities,  FIR, radio continuum (1.4\,GHz), H$\alpha$, $B$-band, and UV (1550\AA\,--\,1650\AA) luminosities for a sample of 23 edge-on late-type spiral galaxies. Typical star formation indicators, such as star formation rates (SFRs), are determined and a statistical multi-parameter/frequency correlation analysis is carried out.}
{We find strong linear correlations, covering at least two orders of magnitude, between star formation indicators and integrated (disk$+$halo) luminosities in all covered wavebands. In addition to the well established $L_{\rm FIR}/L_{\rm 1.4GHz}$ relation, we show new and highly significant linear dependencies between integrated soft X-ray luminosities and FIR, radio continuum, H$\alpha$, $B$-band, and UV luminosities. Moreover, integrated soft X-ray luminosities correlate well with SFRs and the energy input into the ISM by SNe. The same holds if these quantities are plotted against soft halo X-ray luminosities. Only a weak correlation exists between the dust mass of a galaxy and the corresponding X-ray luminosity. Among soft X-ray luminosities, baryonic, and \ion{H}{i} gas masses, no significant correlations are found. There seems to exist a critical input energy by SNe into the ISM or a SFR threshold for multi-phase halos to show up. It is still not clear whether this threshold is a physical one or represents an instrument dependent sensitivity limit.}
{These findings strongly support our previous results that multi-phase gaseous galaxy halos in late-type spiral galaxies are created and maintained by outflowing gas produced in star formation processes in the disk plane. They conflict with the concept of halos being mainly due to infalling gas from the intergalactic medium.}


\titlerunning{The multi-phase gaseous  halos of star forming late-type galaxies}
\maketitle
%

\section{Introduction}
In Paper\,I \citep{Tu05} we established the morphological correlation and spatial coincidence between gaseous halos in late-type spiral galaxies, traced by their radio continuum, H$\alpha$, and soft diffuse X-ray emission. We further showed that the continuum emission radiated at UV wavelengths ($\sim 210$\,nm) originated in the disks of the galaxies and is well associated with diffuse ionized gas (DIG) in the halo. 
It is generally accepted that the non-thermal radio continuum is a reliable indicator for cosmic ray (CR) halos, that extraplanar soft X-rays are a good tracer of the hot ionized medium (HIM, e.g., produced by supernova remnants (SNRs), in superbubbles or in superwinds), and that the presence of extraplanar DIG is indicative of the so called disk-halo interaction \citep[see][for a recent review]{De04}. Therefore, we concluded that multi-phase gaseous halos are created by star formation (SF) related processes in the disk plane. Our results presented in Paper\,I did not yield convincing evidence that galaxy halos form by infalling gas from an external reservoir as considered e.g. by \citet{Be00}, \citet{To02} or \citet{Pe05}.

Apart from these phenomenological analogies there also should exist strong statistical correlations between total (disk$+$halo) soft X-ray, H$\alpha$ (DIG), and radio continuum luminosities. If this hypothesis is correct, these luminosities should also strongly correlate with other star formation indicators, such as the FIR, $B$-band, and UV luminosity \citep[e.g.,][]{Co92,Re01}, but also with direct tracers of massive star formation, such as star formation rates \citep[SFRs, see][]{ken98}, supernova (SN) energy input rates normalized to the area of active star formation $\dot{E}_{\rm A}^{\rm tot}$ \citep[e.g.,][simply called energy input rates, hereafter]{Da95,Ro03}, and $L_{\rm FIR}/D_{25}^{2}$ ratios \citep{lehe}. 

In the following we investigate the interrelations among star formation related key parameters and luminosities in order to establish significant (statistical) correlations of the up to now purely morphological relations. 

Integrated luminosities are used to test whether the diffuse emission from the disk and halo is coupled to SF, whereas the pure extraplanar soft X-ray emission is put into relation to the above mentioned parameters in order to show that multi-phase halos are indeed tightly correlated with SF processes in the disk.

Finally, we discuss the question of the existence of an energy threshold provided by stellar feedback processes which needs to be exceeded in order to create multi-phase halos. 
From the above mentioned correlations first empirical estimates of the suspected energy threshold are derived.   

\section{Statistical analysis of key parameters}

\subsection{Key parameters and sample selection}
It is straightforward to include additional quantities which trace massive stars in the disk and the energy they provide for the interstellar medium (ISM) and to search for trends among all those parameters. Such supplemental key quantities considered here are the $B$-band and UV luminosites as well as SFRs (or the FIR SFR per unit area, expressed by $L_{\rm FIR}/D^2_{25}$) and energy input rates by SNe, represented by $\dot{E}_{\rm A}^{\rm tot}$. 

We also include variables which are not directly coupled to SF, but trace different neutral phases of the ISM, such as the dust mass and the \ion{H}{i} mass of a galaxy. 

Moreover, the extent of the SN-driven gas and therefore the size of the multi-phase halo also depends on the gravitational potential of a galaxy \citep[e.g.,][]{MF99}. As a first order approximation of this quantity the baryonic mass of a galaxy (stars and \ion{H}{i}) determined from the $K$-band Tully-Fisher relation \citep{bj01} is also included. 

As was already pointed out in Paper\,I, our initial sample is biased against starburst galaxies, that is towards higher SFRs or energy input rates $\dot{E}_{\rm A}^{\rm tot}$. In order to investigate the possible correlations, non-starburst (but actively star forming) galaxies need to be included which have to cover the intermediate and lower energetic ends of these parameters.  
This obstacle was overcome by selecting edge-on ($i>70\degr$) late-type spiral galaxies from the literature for which the above mentioned luminosities have been published and the SF related parameters could be calculated.  In order to reach the largest possible coverage of the parameter space no constraints were imposed on the integrated luminosities, the $S60/S100$ and $L_{\rm FIR}/D^2_{25}$ ratios, and the energy input rates $\dot{E}_{\rm A}^{\rm tot}$. 

Galaxies are considered to have multi-phase gaseous halos, if they show kpc-sized extraplanar radio continuum, DIG, and X-ray emission which surrounds a large fraction of the disk plane. Extended X-shaped DIG structures, as visible on H$\alpha$ images of NGC\,1482 \citep{St04a} or NGC\,5775 \citep{Tu05}, trace most likely a limb brightened outflow cone and are also classified to possess a halo. Single plumes or filamentary structures are not considered to constitute a halo.

It should be kept in mind that the sample is limited in sensitivity and is not complete.

The enlarged sample consists now of 23 galaxies, nine from \citet{Tu05}, seven from \citet[][three of their targets are also in our sample and their results are consistent with ours]{St04a}, and another seven galaxies taken from the literature. The number of targets is sufficiently large to test the individual correlations by means of a statistical analysis. All physical parameters adopted throughout this study are listed for the whole sample in Table~
1.
                                                 
Integrated X-ray luminosities for the first nine galaxies listed in Table~1 are from Paper\,I, the next seven are taken from \citet{St04a}, whereas the last seven stem from other literature sources (see notes to table).
\begin{table*}
\addtocounter{table}{1}
\centering
\caption{Fit parameters and correlation coefficients. Listed are the Spearman rank-order correlation coefficient $r_{\rm s}$, its corresponding significance $t_{\rm s}$, and the p-value of r$_{\rm s}$, together with results derived from linear regression ($Y=mX+b$) for the pairs of parameters plotted in Figs.~\ref{F1}--\ref{F3}. Uncertainties of the gradient $m$ and the intercept $b$ are given on a $1\sigma$ level. $b'$ is the intercept if a slope of $m\,=\,1$ is assumed.} 
\begin{tabular}{llccccccr}
\hline\hline
\noalign{\smallskip}
$\log(Y)$   & $\log(X)$              & m             & b              & b$'$    & red. $\chi^2$ & $r_{\rm s}$ & $t_{\rm s}$ & p-value     \\
\noalign{\smallskip}                                                                                                     
\hline                                                                                                                   
\noalign{\smallskip}                                                                                                     
$L_{\rm FIR}$ & $L_{\rm 1.4GHz}$     & 0.94$\pm$0.06 & +0.85$\pm$0.04 &  +0.86  & 0.030         & 0.97        & 17.4        & $<$0.0001   \\
$L_{\rm X}$ & $L_{\rm 1.4GHz}$       & 0.65$\pm$0.11 & $-0.10\pm$0.11 & $-0.01$ & 0.266         & 0.74        & 4.80        & 0.0001      \\
$L_{\rm X}$ & $L_{\rm FIR}$          & 0.80$\pm$0.14 & $-0.87\pm$0.13 & $-0.99$ & 0.243         & 0.76        & 5.10        & 0.0001      \\
$L_{\rm X}$ & $L_{\rm H\alpha}$      & 0.88$\pm$0.27 & $-0.95\pm$0.22 & $-1.03$ & 0.370         & 0.67        & 3.93        & 0.0009      \\
$L_{\rm X}$ & $L_{\rm B}$            & 1.29$\pm$0.29 & $-1.12\pm$0.19 & $-0.95$ & 0.295         & 0.71        & 4.39        & 0.0003      \\
$L_{\rm X}$ & $L_{\rm UV}$           & 0.87$\pm$0.22 & +0.28$\pm$0.24 &  +0.40  & 0.292         & 0.73        & 4.66        & 0.0002      \\
$L_{\rm X}$ & $SFR_{\rm FIR}$        & 0.79$\pm$0.15 & $-0.58\pm$0.11 & $-0.64$ & 0.252         & 0.76        & 5.10        & 0.0001      \\
$L_{\rm X}$ & $SFR_{\rm H\alpha}$    & 0.86$\pm$0.27 & $-0.07\pm$0.18 & $-0.02$ & 0.378         & 0.64        & 3.63        & 0.0018      \\
$L_{\rm X}$ & $L_{\rm FIR}/D^2_{25}$ & 0.73$\pm$0.17 & $-1.07\pm$0.19 & $-1.32$ & 0.303         & 0.69        & 4.16        & 0.0005      \\
$L_{\rm X}$ & $\dot{E}_{\rm A}$      & 0.68$\pm$0.12 & $-0.47\pm$0.10 & $-0.51$ & 0.230         & 0.78        & 5.43        & $<$0.0001   \\
$L_{\rm X}$ & $M_{\rm dust}$         & 0.70$\pm$0.21 & $-0.27\pm$0.13 & $-0.22$ & 0.386         & 0.58        & 3.10        & 0.0058      \\
$L_{\rm X}$ & $M_{\rm \ion{H}{i}}$   & $0.15\pm$0.33 & $-0.31\pm$0.22 &  +0.09  & 0.578         & 0.09        & 0.39        & 0.6980      \\
$L_{\rm X}$ & $M_{\rm Baryon}$       & $0.51\pm$0.37 & $-0.66\pm$0.25 & $-0.93$ & 0.535         & 0.29        & 1.32        & 0.2022      \\
\noalign{\smallskip}
\hline
\noalign{\smallskip}                                                                                                     
$L_{\rm X,h}$ & $SFR_{\rm FIR}$        & 0.93$\pm$0.17 & $-1.13\pm$0.14 & $-1.11$ & 0.189         & 0.72        & 3.42        & 0.0057     \\
$L_{\rm X,h}$ & $SFR_{\rm H\alpha}$    & 1.13$\pm$0.22 & $-0.31\pm$0.15 & $-0.35$ & 0.203         & 0.71        & 3.26        & 0.0075     \\
$L_{\rm X,h}$ & $\dot{E}_{\rm A}$      & 0.77$\pm$0.21 & $-1.12\pm$0.19 & $-1.14$ & 0.308         & 0.69        & 2.96        & 0.0144     \\
$L_{\rm X,h}$ & $L_{\rm FIR}/D^2_{25}$ & 0.70$\pm$0.32 & $-1.60\pm$0.45 & $-1.89$ & 0.449         & 0.61        & 2.45        & 0.0341     \\
\noalign{\smallskip}
\hline
\label{tab2}
\end{tabular}
\end{table*}

\subsection{Statistical analysis}
We performed least-squares fitting (assuming $Y=mX+b$) and Spearman rank-order correlation analysis, to test the significance of the correlation between the investigated pairs of parameters. As these quantities, for example $L_{\rm X}$ and $L_{\rm FIR}$, usually vary only within a certain physical range, e.g., set by the number of stars in the disk, these variables are most likely not normally distributed and their interrelation needs to be investigated by means of rank-order correlation analysis.  

The slopes $m$, their intercepts $b$ and $b'$ (for $m=1.0$), the reduced $\chi^2$, the Spearman rank-order correlation coefficients $r_{\rm s}$, $t_{\rm s}$, the levels of significance of $r_{\rm s}$, and the corresponding p-values of $r_{\rm s}$ are listed in Table~\ref{tab2}. In the case that both quantities are not linearly correlated (null hypothesis), $t_{\rm s}$ follows a Student's $t$-distribution with $n-2$ degrees of freedom. The null hypothesis turns out to be wrong if $|t_{\rm s}| >t_{\rm n-2,1-\alpha}$, where $\alpha$ is the level of significance, which we adopt to be 0.01. The quantile $t_{\rm 21,0.99}$ of the $t$-distribution, as tabulated in relevant statistics textbooks \citep[e.g.,][]{Ha}, is 2.518. For all $t_{\rm s}$-values which are larger, the null hypothesis needs to be discarded (with a $99\%$ confidence level) which means that all those pairs of parameters are strongly correlated.

Moreover, if the p-values listed in Table~\ref{tab2} for the integrated luminosities are less than the adopted $\alpha$-level, the null hypothesis needs to be discarded, too. In the present case only \ion{H}{i} and baryonic masses do not fulfill these criteria and are therefore assumed not to correlate with soft X-ray luminosities. 

The same analysis was carried out for the extraplanar soft X-ray emission ($L_{\rm X,h}$) and the derived parameters are also listed in Table~\ref{tab2}. All correlations and their corresponding least-square fits are shown in Fig.~\ref{F3}. 

\begin{figure*}
\centering
\includegraphics[width=15.75cm,height=15.5cm,angle=-90]{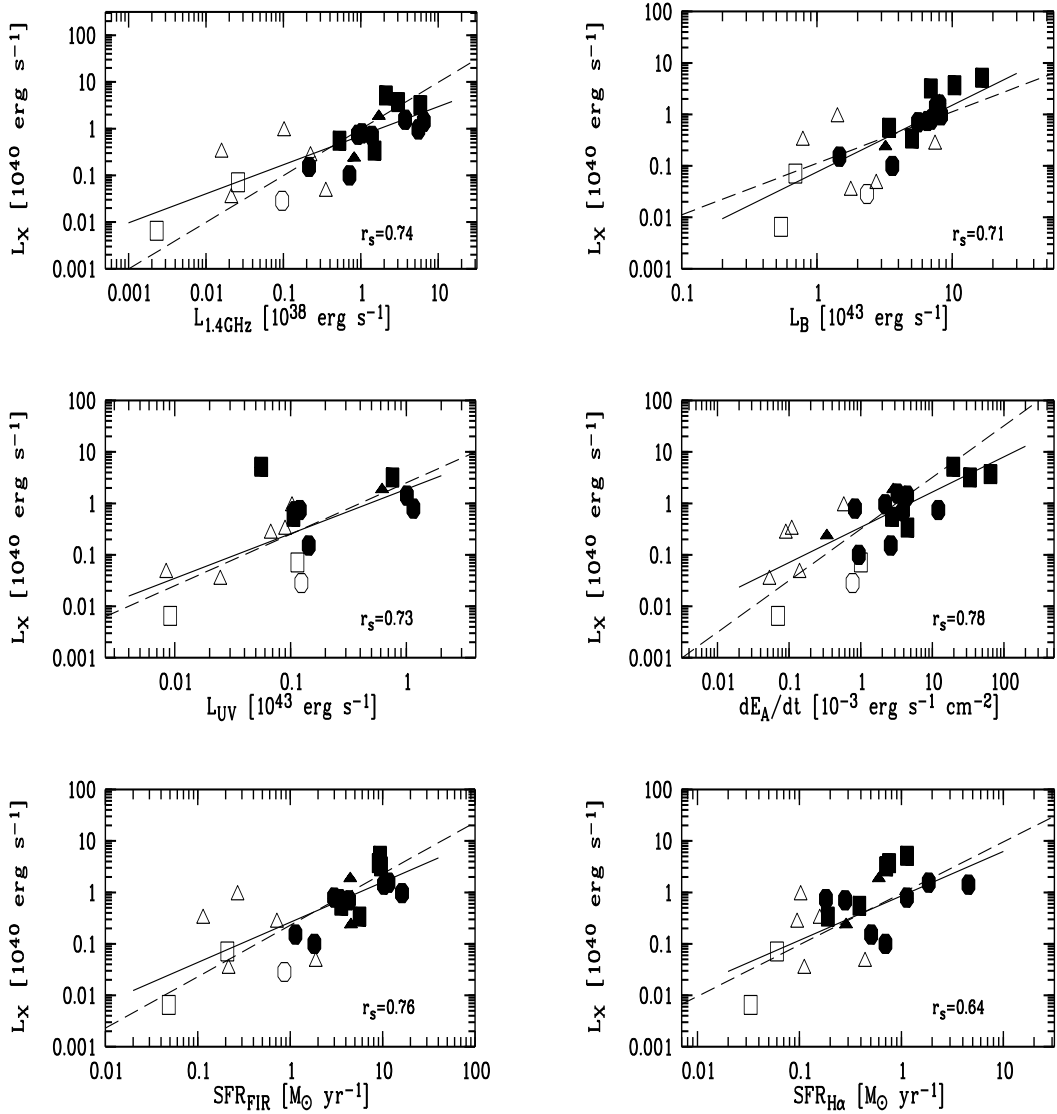}
\caption{Diagnostic diagrams confirming the strong correlations between integrated radio continuum, $B$-band, UV, and soft X-ray (0.3\,--\,2.0\,keV) luminosities. X-ray luminosities also correlate well with the energy input rate by SNe ($\dot{E}_{\rm A}$) and with FIR and H$\alpha$ SFRs. Circles address our sample, squares represent data from \citet{St04b}, and triangles denote data, collected from the literature (cf. Table.~1). Filled (open) symbols refer to galaxies with (un)detected multi-phase gaseous halos. Solid lines are best fits from linear regression while dashed lines indicate the trend expected for a relationship of unit slope. $r_{\rm s}$ is the Spearman rank-order correlation coefficient.}
\label{F1} 
\end{figure*}

\begin{figure*}
\centering
\includegraphics[width=10.5cm,height=15.5cm,angle=-90]{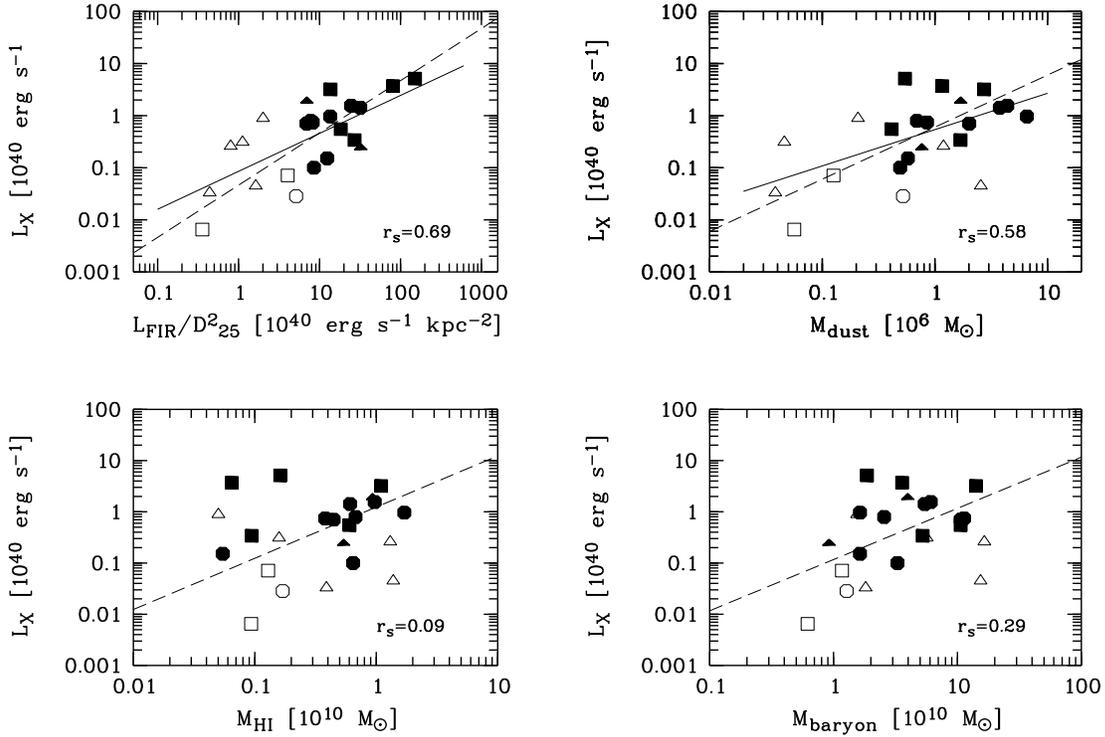}
\caption{Functional dependence between soft X-ray luminosities and $L_{\rm FIR}/D^{2}_{25}$ (upper left), the dust mass of a galaxy (upper right), its \ion{H}{i} mass (lower left), and the baryonic mass (lower right).} 
\label{F2} 
\end{figure*}

\begin{figure*}
\centering
\includegraphics[width=10.5cm,height=15.5cm,angle=-90]{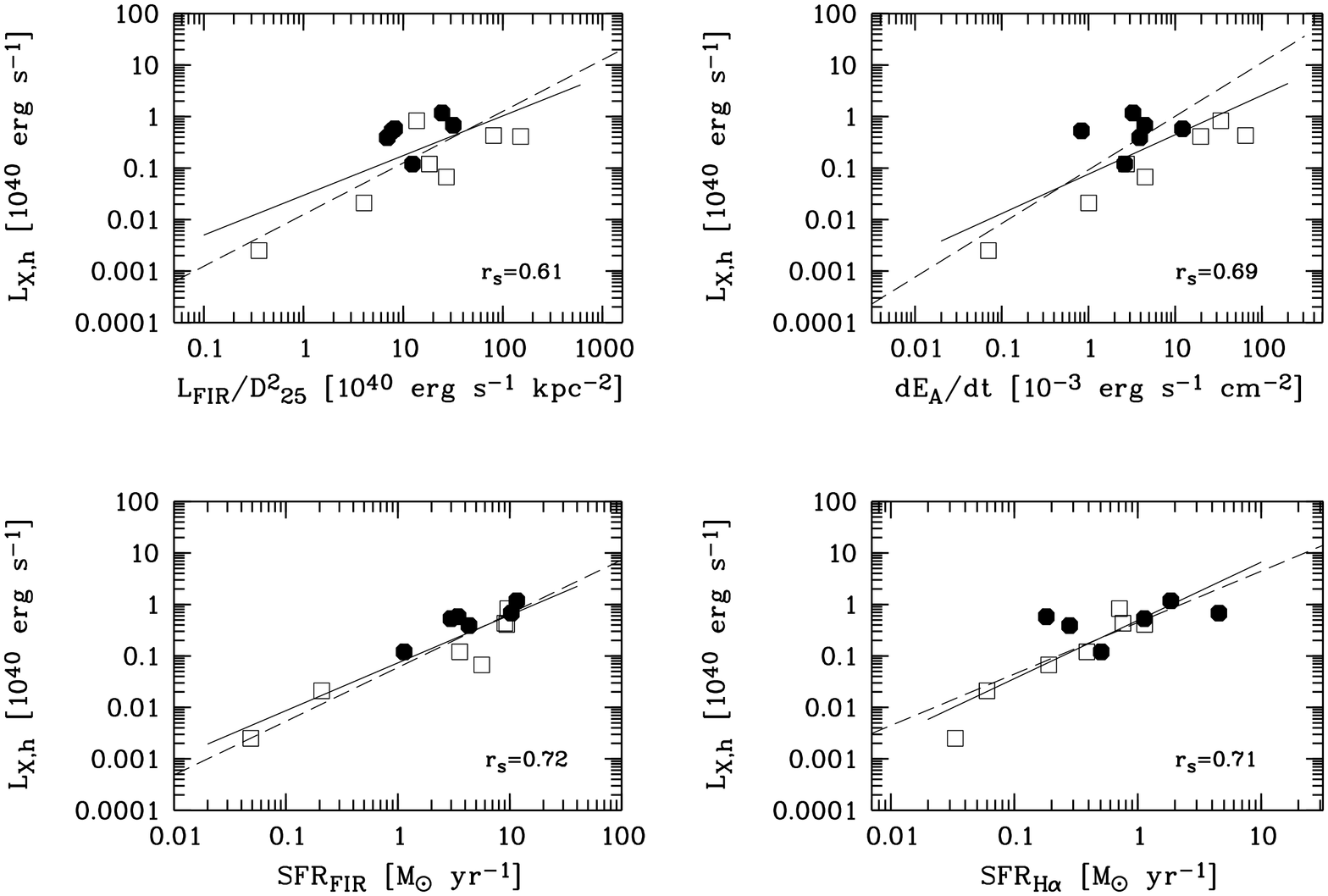}
\caption{Functional dependence between $L_{\rm X,h}$, the soft (0.3\,--\,2.0\,keV) X-ray luminosity of the halo, and $L_{\rm FIR}/D^{2}_{25}$ (upper left), $\dot E_{\rm A}$ (upper right), $SFR_{\rm FIR}$ (lower left), and $SFR_{\rm H\alpha}$ (lower right). Solid symbols represent our sample, open symbols denote data taken from \citet{St04a}. Solid and dashed lines have the same meaning as described in Fig.~\ref{F1}.} 
\label{F3} 
\end{figure*}
\section{Results and Discussion}
\subsection{Correlations with integrated luminosities}
Based upon statistical analysis of the sample strong linear correlations ($r_{\rm s}$\,$\ge$\,0.67) between star formation indicators and integrated (disk$+$halo) multi-frequency luminosities could be established. In addition to the generally accepted $L_{\rm FIR}/L_{\rm 1.4GHz}$ relation \citep[e.g.,][see also Table~2 of this work]{dej85,Co92}, highly significant linear dependencies between integrated soft X-ray luminosities (0.3\,--\,2.0\,keV) and integrated radio continuum (1.4\,GHz), $B$-band, UV, H$\alpha$, and FIR luminosities have been found (see Fig.~\ref{F1})\footnote{The latter two relations are not shown here, as the H$\alpha$ and FIR luminosities are directly proportional to the corresponding SFRs \citep[see][]{ken98}.} Moreover, the integrated soft X-ray luminosities correlate well with star formation rates and SNe energy input rates (Figs.~\ref{F1} and \ref{F2}). These correlations unambiguously confirm that the diffuse radiation of the X-ray emission and the DIG, as well as the CRs detected in the disks and halos are indeed coupled to SF.

Although H$\alpha$, B, and UV-band luminosities are generally a good tracer of the SF activity, these quantities can be seriously affected by absorption internal to the galaxy. The typical scatter of about 1\,dex visible in our correlations might be indicative of this effect. However, absorption appears to be unlikely to lower the luminosities in such a way that the correlations are seriously weakened or even destroyed.

The strongly discrepant data point visible in the $L_{\rm X}$/$L_{\rm UV}$-diagram represents measurements for one of the most extreme starburst galaxies, M\,82. In order for M\,82 to follow our correlation, a much higher (lower) UV (X-ray) luminosity of about 1\,dex is required. However, independent observations carried out in the UV \citep{CW82,Rif95,Ho05} and at X-ray energies \citep{Fab92,St04a} yield very consistent results which imply that this ``deviation'' is an intrinsic feature of this outstanding galaxy. 

More than 60\% of the UV emission of M\,82 measured at about 1530\AA\ is extraplanar in origin \citep{Ho05}. These authors claim that extraplanar star formation or a combination of photoionization and shock ionization either appear unlikely or cannot account for the halo emission in the UV. Instead stellar continuum radiation from the disk plane scattered at dust particles in the halo shall be the main mechanism causing the UV emission. Given that M\,82 is a low mass spiral galaxy ($M_{\rm bar}=1.86\times10^{10}{\rm M}_{\sun}$, Table~1) dust is assumed to be driven relatively easily by the superwind to high extraplanar distances. This independently supports the ``scattering''-hypothesis and could also explain the lacking UV emission, provided the $(L_{\rm X},L_{\rm UV})$-relation is valid.

Alternatively, the high X-ray and low UV luminosity of M\,82 can be explained by gas within the starburst region which is thermalized beyond the UV excitation limit. In such a extreme case in which the diffuse soft X-ray emission is dominating the energy output, we expect a deviation from the linear correlation curve during phases of strong starburst activity.

Apparently only a weak correlation between the dust mass of a galaxy and the corresponding soft X-ray luminosity exists. Dust, as a byproduct of SF, is correlated with the FIR luminosity and should therefore also correlate (at least weakly) with other SF indicators, such as X-ray luminosities. 

The non-correlation between $L_{\rm X}$ and $M_{\ion{H}{i}}$ is also not surprising as the neutral gas component of the ISM is not directly involved in SF. 

Surprisingly only at a first glance, a correlation between galaxy potential and X-ray luminosities is not found. As the baryonic mass \citep{bj01} is calculated from rotational velocities of the stars and the \ion{H}{i}-gas in a galaxy, this parameter traces the dynamics of the system and is therefore insensitive to SF. Nevertheless, this finding bears two important implications. Firstly, the dynamics of the gas is apparently not related to the diffuse emission produced in SF events. Secondly, for a more accurate estimate of the baryonic mass a better mass-related approximation needs to be found. This relation should also include dark matter halos in order to study the effects this additional component might have on the established relations. 
If there is still no correlation with baryonic mass, then the targets of our sample do not follow the $L_{\rm X} \sim v_{\rm rot}^{5}$ relation derived from the infall simulations by \citet{To02}.

It should be pointed out that $M_{\rm bar}$ represents the gravitational potential of a galaxy and should therefore be sensitive to the overall extent of the multi-phase halo. However, it is not sensitive to the existence/non-existence of such halos. In this regard the SFR and the energy input rate by SNe into the ISM turn out to be the most important parameters.  

\subsection{Correlations with extraplanar luminosities} In order to establish that multi-phase halos are indeed a product of SF, the pure extraplanar emission in the radio continuum, in H$\alpha$, and at soft X-ray energies also needs to correlate with SFRs and the energy input rate by SNe. 

At present this analysis can best be done in the X-ray regime. Other wavelengths, such as H$\alpha$ and the radio continuum, are not usable, because the integral disk and halo luminosities do not allow to reliably estimate the fraction of the luminosity originating in the halo. For the DIG, this fraction can vary from 12\% to 60\% \citep{Ro00} and would therefore lead to unacceptable uncertainties.

From the original sample, 13 galaxies are suitable for the ``halo''-sample (those from \citet{St04a} and \citet{Tu05}). The remaining galaxies either do not show extraplanar X-ray emission or the published data did not allow a disentangling of disk and halo luminosities. 

It should be pointed out that inclination issues, regarding the reliability of detecting multi-phase halos, are negligible. Either because extraplanar gas was detected in case of inclination angles $\le 80\degr$ \citep[see NGC\,1482,][]{St04a}, or because the inclination-corrected extent of the ``extraplanar'' gas was indeed not large enough to be seen above the disk plane \citep[see NGC\,3877,][]{Tu05}. Most importantly, $L_{\rm X}$ and $L_{\rm X,h}$ are not linearly correlated with the inclination angle $i$. This is proved by low Spearman rank-order correlation coefficients of $r_{\rm s}=-0.26$ and $r_{\rm s}=0.23$, respectively, again on a $99\%$ confidence level.

In Fig.~\ref{F3} we plot the soft (0.3\,--\,2.0\,keV) halo luminosity ($L_{\rm X,h}$) as a function of the different SF indicators (which are $SFR_{\rm FIR}$, $SFR_{\rm H\alpha}$, $L_{\rm FIR}/D^{2}_{25}$, and $\dot{E}_{\rm A}$). All pairs of parameters show strong correlations ($r_{\rm s}\ge 0.67$) on a $99\%$ confidence level, except $L_{\rm X,h}$ vs. $L_{\rm FIR}/D^{2}_{25}$ ($r_{\rm s}=0.61$). A possible explanation could be that $L_{\rm FIR}/D^{2}_{25}$, a measure of the SF rate intensity per unit disk area, underestimates the real SF intensity. In disk galaxies SF usually occurs within the H$\alpha$-disk and is not that widespread as the distribution of the old stellar population (measured by $D_{25}$) implies. 

It turns out that the FIR and the H$\alpha$ SFRs are the most important parameters which decide on the formation of X-ray halos. Among these two quantities, $SFR_{\rm FIR}$ seems to be the more reliable parameter, because it is very little affected by dust absorption and is dominated by disk emission. On the other hand, the H$\alpha$ emisson, and therefore $SFR_{\rm H\alpha}$, is a more sensitive measure of how much gas is converted into massive stars. However, this quantity is significantly affected by extinction and in some cases most likely dominated by extraplanar emission. 

Given the strong correlations between integrated X-ray, H$\alpha$, and radio continuum luminosities as well as between halo X-ray luminosities and SFRs or $\dot{E}_{\rm A}$, it is plausible, to adopt similar relations for the extraplanar DIG and the CR halos. Therefore, we expect the SFR and the energy input rate by SNe to be the main driving-agents for multi-phase galaxy halos. Our statistical results are also fully consistent with the morphological and spatial coincidences presented in Paper\,I. 
For consistency reasons it is certainly worthwhile to re-examine existing radio and DIG data to allow for a clear separation of disk and halo luminosities.

\subsection{Constraining the star formation threshold}
In order to achieve a more comprehensive picture on the evolution of galaxy halos in different wave bands, we need to answer the question of the existence of a threshold SFR (or equivalently the SN energy input rate $\dot E_{A}$) required to create multi-phase gaseous halos. The existence of such a threshold can be motivated as follows: It is well known that hot ionized gas is driven off-plane via superbubbles produced by SNe. From a theoretical study of breakout conditions of such superbubbles \citep{LC88} we know that the mechanical luminosity of SNe is a key quantity which determines whether a superbubble will breakout or collapse. In other words, the overpressured X-ray emitting gas will leave the disk and form a halo if the energy provided by SNe is above a certain limit. 

On the other hand, it can be argued that channels, through which the hot gas reaches the halo, were already blown into the ISM during previous periods of intense star formation. Hence, a substantially lower SFR is needed to elevate the gas into the halo, making it even more difficult to establish a threshold by means of observations.

From Fig.~\ref{F1} the existence of a critical SFR (or energy input rate) is hard to constrain, as the lower energy end of our correlations is statistically not well covered. It appears, however, that galaxies without halos (open symbols) also follow a linear relation and that the region dividing galaxies with halos from those without is relatively narrow. For multi-phase halos to evolve, the data imply that a critical threshold of $SFR_{\rm FIR}\ge 1.0$\,{\rm M$_{\odot}$/yr}, $SFR_{\rm H\alpha}\ge 0.1$\,{\rm M$_{\odot}$/yr} or $\dot{E}_{\rm A} \ge 1\times 10^{-3}$\, erg\, s$^{-1}$\,cm$^{-2}$ needs to be exceeded. Interestingly, DIG seems to coexist with other gas components for H$\alpha$ energy fluxes larger than $(3.2\pm0.5)\times10^{40}$\ erg\ s$^{-1}$\,kpc$^{-2}$ \citep{Ro03}. 

Unfortunately, with the present data a detection threshold, introducing systematic uncertainties by the limited sensitivity of the instrument, cannot be a priori ruled out. 
In order to distinguish between a significant physical threshold and a sensitivity limit as well as to reach a better understanding on the formation of gaseous galaxy halos, deep observations of galaxies with no or only little halo emission are required.   
Therefore, future studies should aim at increasing the sample to test the low energy end of our correlations and to investigate the above relations for galaxies of different Hubble-types, such as Sa and Irr, as the different SF histories of these galaxies might lead to different relations and thresholds.       
\section{Summary and Conclusions}
With a sample of 23 actively star forming galaxies studied in the X-ray regime and by implementing additional wave bands (H$\alpha$ and UV), we found remarkably strong linear correlations between integrated 1.4GHz radio continuum, FIR, H$\alpha$, $B$-band, UV, and soft X-ray luminosities. Strong correlations also exist if soft X-ray luminosities are plotted against SFRs, $L_{\rm FIR}/D^{2}_{25}$, or the energy input rate by SNe per unit area, expressed by $\dot E_{\rm A}$. Integrated X-ray luminosities neither correlate with the \ion{H}{i} nor with the baryonic mass of a galaxy. 

Strong correlations are also found if the diffuse soft X-ray luminosity of the {\em halo} is plotted against the FIR and H$\alpha$ SFR or the SN energy input rate $\dot E_{\rm A}$. These quantities are considerd to be the most important parameters for the creation of multi-phase halos. 

If a critical energy threshold exists and an instrumental detection bias is negligible, the present data suggest that a threshold of $SFR_{\rm FIR}\ge 1.0$\,{\rm M$_{\odot}$/yr}, $SFR_{\rm H\alpha}\ge 0.1$\,{\rm M$_{\odot}$/yr} or $\dot{E}_{\rm A} \ge 1\times 10^{-3}$\, erg\, s$^{-1}$\,cm$^{-2}$ needs to be exceeded in order to create multi-phase galaxy halos.

Our results are in agreement with previous findings of a morphological and spatial coincidence of gaseous multi-phase halos (see Paper\,I). They clearly imply that multi-phase halos are the consequence of stellar feedback processes in the disk plane \citep[e.g.,][]{Av04,Av05} but conflict with the concept of halos being due to infalling gas from the intergalactic medium \citep[e.g.,][]{Be00,To02,Pe05}.

\begin{acknowledgements}
RT acknowledges financial support by Deutsches Zentrum f\"ur Luft-- und Raumfahrt (DLR) through grant 50\,OR\,0102. We appreciate the comments pointed out by the anonymous referee.
This work has made use of the SIMBAD database and of HyperLeda (http://leda.univ-lyon1.fr/). 
\end{acknowledgements}

\begin{landscape}
\addtocounter{table}{-2}
\begin{table*}
\hspace{-8cm}
\begin{minipage}[b]{25.25cm}
\begin{center}
\caption{Physical properties of the sample.}
\begin{tabular}{cccccccccccccccc}
\hline\hline
\noalign{\smallskip}
Galaxy  & $d$ & $i$  & $L_{\rm 1.4GHz}$ & $L_{\rm FIR}$ & $L_{\rm B_{T}^{0}}$   & $L_{\rm UV}$  & $L_{\rm H\alpha}$  & $L_{\rm X,soft}$ &$SFR_{\rm FIR}$ & $SFR_{\rm H\alpha}$ & $\dot{E}_{\rm A}^{\rm tot}$ & $L_{\rm FIR}/D_{25}^{2}$& $M_{\rm bar}$ & $M_{\ion{H}{i}}$ & $M_{\rm dust}$  \\ 
\noalign{\smallskip}
 & [Mpc]& $[\degr]$& [$10^{38}\,{\rm erg\ s^{-1}}$] & \multicolumn{3}{c}{[$10^{43}\,{\rm erg\ s^{-1}}$]} & \multicolumn{2}{c}{[$10^{40}\,{\rm erg\ s^{-1}}$]} & \multicolumn{2}{c}{[${\rm M}_{\sun}\ {\rm yr}^{-1}$]} & [$10^{-3}\,{\rm erg\,s^{-1} cm^{-2}}$] & [$10^{40}\,{\rm erg\ s^{-1}\ kpc^{-2}}$] &  \multicolumn{2}{c}{[$10^{10} {\rm M}_{\sun}$]} & [$10^{6} {\rm M}_{\sun}$] \\
\noalign{\smallskip}
(1) & (2)& (3) & (4) & (5) & (6) & (7) & (8) & (9) & \multicolumn{2}{c}{(10)} & (11) & (12) & (13) & (14) & (15) \\ 
\noalign{\smallskip}
\hline
\noalign{\smallskip}
NGC\,0891 &  9.5& 88& 1.19$^{a}$& 9.62&5.59&0.11 &   3.54$^{h}$&  0.70$^{r}$    &4.32& 0.28& 3.90$^{u}$ & 6.92 & 10.6 & 0.45 & 2.02 \\
NGC\,3044 & 17.2& 84& 0.62$^{b}$& 4.06&3.61&---  &   8.83$^{j}$&  0.09$^{r}$    &1.83& 0.70& 0.94$^{g}$ & 8.52 &3.28  & 0.64 & 0.49 \\
NGC\,3221 & 54.8& 77& 3.29$^{b}$& 36.0&8.28&---  &        ---  &  1.07$^{r}$    &16.2& --- & 2.20$^{u}$ & 13.5 & 1.63 & 1.70 & 6.55 \\
NGC\,3628 & 10.0& 80& 0.93$^{b}$& 7.66&6.54&0.12 &   2.30$^{k}$&  0.74$^{r}$    &3.45& 0.18& 12.2$^{u}$ & 8.24 & 11.3 & 0.38 & 0.85 \\
NGC\,3877 & 12.1& 83& 0.10$^{b}$& 1.92&2.34&0.12 &   ---       &  0.02$^{r}$    &0.86& --- & 0.77$^{u}$ & 5.14 & 1.28 & 0.17 & 0.52 \\
NGC\,4631 &  7.5& 86& 1.22$^{b}$& 6.63&7.00&1.15 &   14.3$^{k}$&  0.78$^{r}$    &2.99& 1.13& 0.83$^{g}$ & 7.76 & 2.56 & 0.68 & 0.69 \\
NGC\,4634 & 19.1& 83& 0.21$^{c}$& 2.53&1.46&0.14 &   6.40$^{j}$&  0.15$^{r}$    &1.14& 0.51& 2.62$^{u}$ & 12.4 & 1.63 & 0.05 & 0.58 \\
NGC\,4666 & 20.2& 80& 2.64$^{b}$& 23.0&7.56&1.01 &   57.5$^{d}$&  1.42$^{r}$    &10.3& 4.55& 4.46$^{g}$ & 31.9 & 5.44 & 0.61 & 3.77 \\
NGC\,5775 & 26.7& 84& 2.82$^{b}$& 25.7&8.05&---  &   23.4$^{l}$&  1.55$^{r}$    &11.6& 1.85& 3.26$^{g}$ & 24.6 & 6.09 & 0.97 & 4.39 \\
\noalign{\smallskip}                                           
\hline                                                         
\noalign{\smallskip}                                           
                                                               
M\,82     &  3.6& 82& 2.10$^{b}$& 20.9&16.7& 0.06&   14.4$^{h}$& 5.15$^{k}$     &9.39& 1.14& 19.6$^{\ \,}$ & 151  & 1.86 & 0.16 & 0.54 \\
NGC\,0253 &  2.6& 79& 0.53$^{b}$& 7.93&3.42& 0.11&   4.90$^{h}$& 0.55$^{k}$     &3.57& 0.39& 2.74$^{\ \,}$ & 18.3 & 10.6 & 0.60 & 0.41 \\
NGC\,1482 & 22.1& 58& 3.04$^{b}$& 20.1&10.4& --- &   9.60$^{k}$& 3.70$^{k}$     &9.65& 0.76& 64.9$^{\ \,}$ & 80.7  & 3.58 & 0.06 & 1.16 \\
NGC\,3079 & 17.1& 85& 5.88$^{b}$& 21.3&6.94& 0.76&   9.00$^{k}$& 3.20$^{k}$     &9.60& 0.71& 33.6$^{\ \,}$ & 13.7 & 14.1 & 1.10 & 2.74 \\
NGC\,4244 &  3.6& 85&0.002$^{c}$& 0.11&0.54& 0.01&   0.42$^{k}$& 0.007$^{k}$    &0.05&0.03& 0.05$^{g}$ & 0.36 & 0.62 & 0.09 & 0.06 \\
NGC\,4945 &  3.7& 78& 1.51$^{c}$& 12.5&5.05& --- &   2.40$^{k}$& 0.34$^{k}$     &5.62& 0.19& 4.50$^{u}$ & 27.1 & 5.24 & 0.09 & 1.69 \\
NGC\,6503 &  5.2& 75& 0.03$^{b}$& 0.46&0.69& 0.11&   0.76$^{k}$& 0.07$^{k}$     &0.21& 0.06& 1.00$^{\ \,}$ & 4.04 & 1.17 & 0.13 & 0.13 \\
\noalign{\smallskip}                                           
\hline                                                         
\noalign{\smallskip}                                           
NGC\,0055 & 1.6 & 80& \hspace{0.03cm}0.02$^{a}$& 0.95&0.79& 0.09&   2.00$^{m}$&\hspace{0.07cm}0.31$^{s}$ & 0.12 & 0.16 & 0.11$^{\ }$ & 1.12 & 5.65 & 0.16 & 0.05 \\
NGC\,1511 & 17.5& 72& 0.82$^{f}$& 10.1&3.21& --- &   3.63$^{n}$& 0.26$^{n}$     &4.54& 0.29& 0.34$^{n}$ & 32.4 & 0.92 & 0.54 & 0.77 \\
NGC\,3556 & 14.1& 79& 1.71$^{b}$& 9.92&8.05& 0.62& \hspace{0.1cm}7.60$^{i}$& 2.00$^{t}$     &4.46& 0.60 & 2.85$^{e}$ & 6.96 & 3.97 & 0.93 & 1.70 \\
NGC\,4522 & 16.0& 79& 0.10$^{c}$& 0.60&1.42& 0.10&   1.30$^{o}$& 0.89$^{o}$     &0.27& 0.10& \hspace{0.03cm}0.58$^{\ \,}$ & 2.00 & 1.55 & 0.05 & 0.21 \\
NGC\,4565 & 9.7 & 89& 0.22$^{b}$& 1.60&7.47& 0.07&   1.20$^{p}$& 0.26$^{p}$     &0.72& 0.09& 0.09$^{g}$ & 0.80 & 16.5 & 1.30 & 1.19 \\
NGC\,4656 & 7.5 & 90& 0.02$^{b}$& 0.48&1.78& 0.02&   1.41$^{v}$& 0.03$^{p}$     &0.22& 0.11& 0.05$^{g}$ & 0.44 & 1.81 & 0.39 & 0.04 \\
NGC\,5907 & 14.9& 86& 0.35$^{b}$& 4.32&2.74& 0.01& \hspace{0.035cm}5.57$^{q}$& 0.05$^{p}$     &1.89& 0.44& 0.14$^{g}$ & 1.63 & 15.4 & 1.38 & 2.56 \\
\noalign{\smallskip}
\hline
\end{tabular}
\end{center}
\vspace{-0.2cm}
{\small Notes:\quad Cols. (1) to (3): Name, distance, and inclination of the target. Col. (4): Radio continuum luminosities measured at 1.49\,GHz. Col. (5): FIR-luminosities calculated from revised IRAS $S_{60}$ and $S_{100}$ flux densities \citep{Sa03}. Col. (6): Total $B_{\rm {T}}^{\rm 0}$ luminosities calculated from RC3 \citep{rc3} and corrected for redshift, Galactic and internal extinction, assuming reddening values from \citep{BH}. Col. (7): De-reddened UV-continuum luminosities measured at 1650\AA\ \citep{Rif95} and 1550\AA\ \citep{CW82}, respectively. The UV flux for NGC\,4656 has been determined at 2270\AA\ \citep{Ma96}. No correction for redshift has been applied. Col. (8): H$\alpha+[\ion{N}{ii}]$-luminosities, uncorrected for extinction and [\ion{N}{ii}]-contamination. Col. (9): Integrated (disk+halo) diffuse X-ray luminosities obtained from spectral fitting between 0.3 and 2.0\,keV. For NGC\,3221 and NGC\,3877 disk only. Col. (10): Total star formation rate, based on FIR and H$\alpha$ luminosities according to \citep{ken98}. Col. (11): SN energy input rates per unit area \citep{Da95,Co92} for M\,$\,>$\,5M$_{\sun}$. Col. (12): The optical diameter of the 25$^{\rm th}$ magnitude isophote ($D_{25}$) was taken from RC3 \citep{rc3}. Col. (13): Baryonic masses were calculated according to \citet{bj01}. Col. (14): To derive the \ion{H}{i} gas mass, we used $M_{\ion{H}{i}}=4.78\times 10^5\,d^2\,S(\ion{H}{i})$, where $S(\ion{H}{i})$, the total \ion{H}{i} flux in units of [Jy\,km/s], has been taken from \citet{HR89}.  Col. (15): Dust masses were calculated following \citet{Hb83}. \\
References:\quad  {\small $(a)$} \citet{Co96}; {\small $(b)$} \citet{Co90}; {\small $(c)$} \citet{Co98}; {\small $(d)$} \citet{Da97}; {\small $(e)$} \citet{Ir99}; {\small $(f)$} \citet{Da01}; {\small $(g)$} \citet{Da95}; {\small $(h)$} \citet{Bell03}; {\small $(i)$} \citet{Yo96}; {\small $(j)$} \citet{Ro00}; {\small $(k)$} \citet{St04a}; {\small $(l)$} \citep{lehe}; {\small $(m)$} \citep{Fe96}; {\small $(n)$} \citet{Da03}; {\small $(o)$} \citet{keko}; {\small $(p)$} \citet{Vo96}; {\small $(q)$} Rand (priv. com.); {\small $(r)$} \citet{Tu05}; {\small $(s)$} \citet{schlegel}; {\small $(t)$} \citet{Wa03}; {\small $(u)$} \citet{Ro03}; {\small $(v)$} Aldering (priv. com.). }
\label{tab1}
\end{minipage}
\end{table*}
\end{landscape}
\end{document}